# Lattice dynamics of KAgF$_3$ perovskite, unique 1D antiferromagnet

Kacper Koteras [1,*], Jakub Gawraczyński [1,*], Mariana Derzsi [1,2], Zoran Mazej [3] and Wojciech Grochala [1,*]

[1] Centre of New Technologies, University of Warsaw, S. Banacha 2C, 02-097 Warsaw, Poland
[2] Advanced Technologies Research Institute, Faculty of Materials Science and Technology in Trnava, Slovak University of Technology in Bratislava, Jána Bottu 8857/25, 917 24 Trnava, Slovakia
[3] Jožef Stefan Institute, Jamova cesta 39, 1000 Ljubljana, Slovenia

* Correspondence: k.koteras@cent.uw.edu.pl (K.K.), j.gawraczynski@cent.uw.edu.pl (J.G.), w.grochala@cent.uw.edu.pl (W.G.)

**Abstract:** Theoretical DFT calculations using GGA+U and HSE06 frameworks enabled vibrational mode assignment and partial (atomic) phonon DOS determination in KAgF$_3$ perovskite, a low-dimensional magnetic fluoroargentate(II). Twelve bands in the spectra of KAgF$_3$ were assigned to either IR active or Raman active modes, reaching very good correlation with experimental values ($R^2$>0.997). Low-temperature Raman measurements indicate that the intriguing spin-Peierls-like phase transition at 230 K is an order-disorder transition and it does not strongly impact the vibrational structure of the material.

*This work is dedicated to Josef Michl at his 80$^{th}$ birthday*

## 1. Introduction

Fluoroargentates(II) attract interest because of their noticeable similarities with isoelectronic copper(II) oxides; many interesting physical phenomena occur in the latter. One of them is low dimensional magnetism, which plays an important role in superconducting materials. KAgF$_3$, which adopts a distorted perovskite structure [1], is an example of such low-dimensional magnetic system. Magnetic susceptibility measurements indicate substantial magnetic anisotropy in this compound, with strong antiferromagnetic ordering along the crystallographic ***b*** axis, and weak ferromagnetic one in the ***ac*** plane (*Pnma* setting).

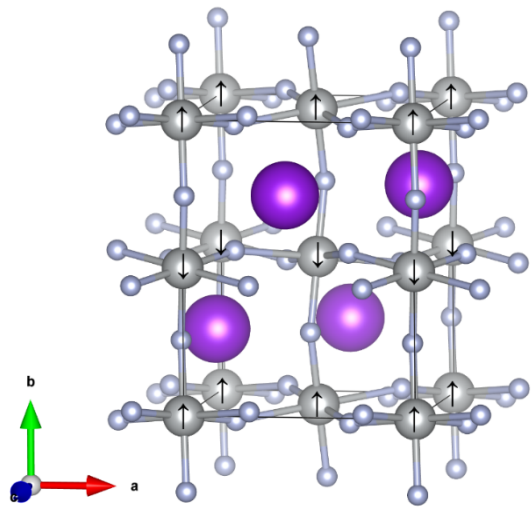

**Figure 1.** Ground state spin ordering in KAgF$_3$ *Pnma* structure as marked by black arrow.

KAgF$_3$ can be regarded as quasi-1D antiferromagnet in which superexchange constant along kinked Ag–F–Ag chains, J$_{1D}$, is about 20 times larger than in the remaining two dimensions [2,3]. Importantly, standard Generalized Gradient Approximation (GGA) approach fails to reproduce antiferromagnetic ground state of KAgF$_3$. Due to considerable correlation of silver valence electrons on *d* orbitals it is necessary to introduce Mott-Hubbard correction within the GGA+U framework [2]. The GGA+U approach permits to reproduce experimental J values for both strong intra-chain and much weaker intra-sheet interactions [3,4]. The unique character of KAgF$_3$ stems from the fact that the intra-chain antiferromagnetic interactions are immensely strong, of the order of 100 meV, thus comparable to those for copper(II) oxides, as well as from substantial magnetic anisotropy.

Interestingly, KAgF$_3$ exhibits an intriguing phase transition at 230 K which resembles a spin-Peierls transition; it has been preliminarily assigned to structural order/disorder-type (*Pnma*/*Pcma*) transition based on X-ray diffraction studies [2]. Moreover, this compound shows a complex magnetic ordering at temperatures below 66 K [4] which is a subject to ongoing investigation. Analysis of X-ray diffraction experiments suggests that phase transition present at ca. 230 K might be of order/disorder type. The low temperature ordered polymorph crystallizes in *Pnma* space group while the high-temperature disordered one in the *Pcma* one, exhibiting tilting of [AgF$_4$]$^{2-}$ units around the *a* and *c* vectors [2]. Most characteristic structural features (tilting of octahedra, Jahn-Teller effect), as well as magnetic features (antiferromagnetic 1D character) are retained during this transition.

Lattice dynamics studies of related AAgF$_3$ systems have been scarce despite the fact that they host interesting physical phenomena. To this date there are no spectroscopic data published on KAgF$_3$. The goal of the current work is to gain insight into lattice dynamics (phonons) of KAgF$_3$ with both experimental and theoretical tools, assign the bands appearing in the Raman scattering and infrared absorption spectra, as well as further elucidate the nature of the 230 K phase transition. Understanding the impact of disorder on the lattice dynamics is also of interest here.

## 2. Experimental

KAgF$_3$ has been prepared using published synthesis route yielding high purity product (equimolar mixture of KF and AgF$_2$ was heated in nickel reactor, under F$_2$ atmosphere, for four days at 300 °C) [2]. Infra-red measurements were carried out on Bruker Vertex 80V vacuum spectrometer using powdered samples placed on HDPE windows. Raman spectra were obtained on Horiba Jobin Yvon LabRam-HR Raman micro-spectrometer with 632.8 nm He–Ne laser exciting beam. As with other Ag$^{II}$ compounds, used laser beam had to be of very low power, in order to minimize thermal (or photo-) decomposition [5,6]. In this work we have used power of circa 0.3 mW. KAgF$_3$ can also easily decompose in contact with the atmosphere, so the sample was enclosed inside a sealed quartz capillary. In case of low-temperature Raman measurements, the capillary was placed in a home-made flow cryostat, described in more detail below, and measured using Horiba Jobin Yvon T64000 Raman spectrometer with 514.5 nm Ar-Kr laser exciting beam, with on-sample power lower than 0.5 mW.

Low-temperature setup was prepared in the following way: 5N Ar gas bottle was connected to FEP coil placed in a 3 L dewar (Figure 2). The amount of LN2 in the dewar and the flow rate of cool Ar gas were chosen to ensure a near constant temperature in the vicinity of the sample (a maximum of 1-2 K drift was observed during measurements).

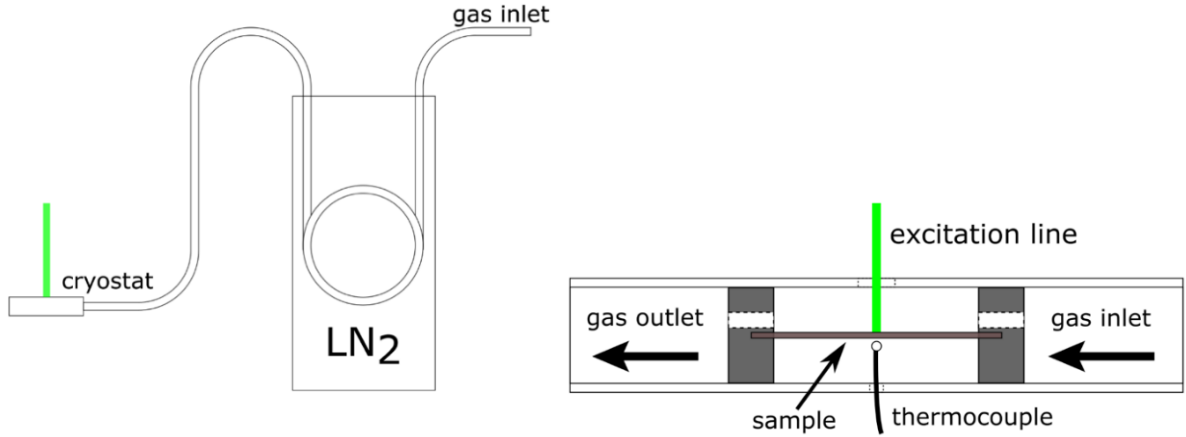

**Figure 2.** The experimental setup for study of Raman spectrum of KAgF$_3$ at low temperature.

The cryostat itself was made from FEP tube with aluminum disks. Holes drilled in the aluminum disks enabled gas flow through the measurement chamber, and retained sample in place. The main body of the cryostat, made from FEP tube, had two holes drilled in on opposing sides of the tube – the first one, about 3 mm in diameter to let in the excitation line, the second one, about 1 mm in diameter, to let in a thermocouple. The temperature inside the cryostat was measured using a thermocouple positioned about 1 mm from the capillary with enclosed sample. To ensure that no condensation formed in the hole that excitation line was coming in, a heat gun was used to blow the newly formed ice crystals away from the cryostat. Minimum stable temperature reached in this setup was equal to 123 K, way below the 230 K order/disorder-type phase transition but above the temperature of magnetic ordering (66 K), and we report here spectra measured at 123 K as well as at room temperature.

## 3. Computational details

Lattice dynamics was modelled using GGA+U framework using Lichtenstein formalism [7] with values of U = 5.0 eV and J = 1.0 eV for silver atoms in order to account for strong electron correlation. Perdew-Burke-Ernzerhof functional revised for solids (PBEsol) [8] and projector-augmented-wave method [9,10] was used, as implemented in VASP 5.2.12 code [11–13]. The cut-off energy of the plane wave basis set was equal to 800 eV with a self-consistent-field convergence criterion of $1 \cdot 10^{-7}$ eV. The k-point mesh spacing used in calculations was equal to 0.16 Å$^{-1}$. Ordered, low temperature, *Pnma* structure of KAgF$_3$ was chosen as a starting model for the calculations [2]. Antiferromagnetic model was assumed in agreement with Zhang *et al.* (AFM along the chains, FM in planes) [3].

Phonon dispersion curves and phonon density of states were calculated for 2 x 2 x 2 supercell with the use of PHONON software package [14,15]. Hellmann-Feynman forces were evaluated using series of atomic displacements (0.0018–0.0025 Å) for which single point energies were calculated with settings mentioned above and harmonic approximation. Furthermore, obtained Γ-point frequencies were compared with values obtained using HSE06 functional [16]. For that comparison the finite differences method was used, as implemented in VASP core. These calculations were carried for KAgF$_3$ unit cell, using less dense k-point mesh spacing of 0.25 Å$^{-1}$ to reduce the calculation cost. This implementation yields frequencies in the centre of the first Brillouin zone (Γ point).

## 3. Results and discussion

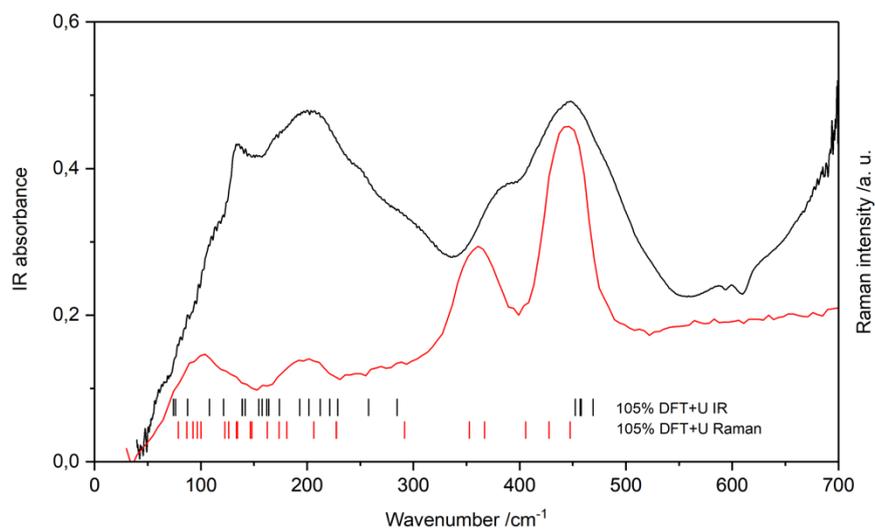

**Figure 3.** FIR (black) and Raman (red) spectra of KAgF$_3$ at room temperature together with marks indicating the computed wavenumbers.

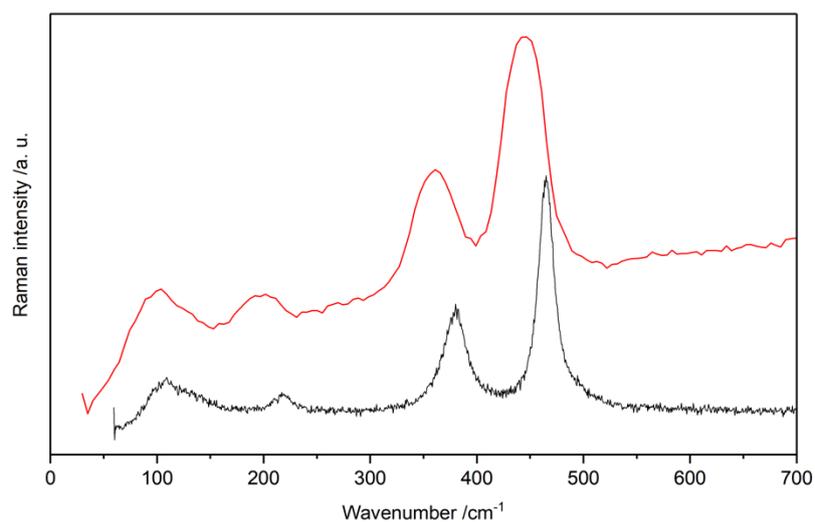

**Figure 4.** Raman spectrum measured at room temperature (red) and 123 K (black).

Infrared and Raman spectra were measured from instrument lower limit up to 1000 cm$^{-1}$. FIR measurement resulted in 8 separate bands, while Raman measurement showed 7 bands. Low temperature Raman measurement exhibits slight temperature shift towards higher wavenumbers. Density functional framework was used to gain in-depth understanding of present spectral features.

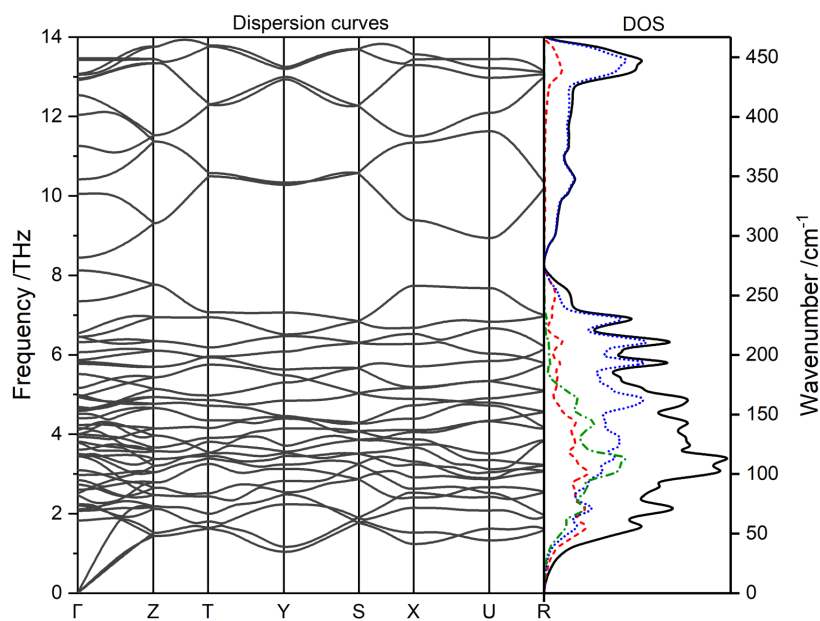

**Figure 5.** The calculated phonon dispersion and phonon (atomic) density of states for KAgF$_3$ (blue dotted line – F, red dashed line – Ag, green dash-dotted line – K)

To set the grounds for analysis of the experimental spectra, we begin discussion with group theory analysis. According to Group theory analysis there are 60 normal vibrational modes for KAgF$_3$ in *Pnma* symmetry ($\Gamma_{acoustic}$ = B$_{1u}$ + B$_{2u}$ + B$_{3u}$; $\Gamma_{optic}$ = 7A$_g$ + 8A$_u$ + 5B$_{1g}$ + 9B$_{1u}$ + 7B$_{2g}$ + 7B$_{2u}$ + 5B$_{3g}$ + 9B$_{3u}$).

Calculated phonon dispersion curves and phonon density of states are shown in Figure 5. No imaginary modes are seen; hence, the low-temperature structure is confirmed to be dynamically stable. Partial (atomic) phonon density of states (Figure 5, right) yields information about atoms involved in oscillations. Up to *ca.* 175 cm$^{-1}$ all atoms contribute comparably to normal modes. The characteristic region of vibrations of potassium atoms is within the energy range 0 – 225 cm$^{-1}$. Silver atoms contribute within slightly broader energy range 0 – 260 cm$^{-1}$ as well as to the highest energy modes between 400 and 450 cm$^{-1}$. Fluorine atoms contribute considerably to all phonon modes of the KAgF$_3$. Their contribution dominates the phonon DOS within the energy range 150 – 250 cm$^{-1}$, which is the characteristic energy region of the F-Ag-F bending vibrations, and the entire higher energy phonon DOS in the range 300 – 450 cm$^{-1}$, which is the characteristic region of the Ag-F stretching vibrations.

The phonon bands belonging to F-Ag-F bending and low-energy lattice modes (below 250 cm$^{-1}$) show relatively low dispersion. On the other hand, phonon bands stemming from diverse Ag-F stretching modes exhibit substantial dispersion indicating considerable coupling of the stretching vibrations along both the direction of the propagation of the AFM chains (corresponding to Y point in the 1$^{st}$ Brillouin zone) and within the [AgF$_2$] planes (X and Z points in the 1$^{st}$ Brillouin zone).

The eigenvectors for selected normal modes assigned to measured IR and Raman spectra are shown in Figure 6, while the complete list of theoretically predicted $\Gamma$-point frequencies is contained in Electronic Supplementary Material.

The measured room-temperature far infra-red absorption spectrum of KAgF$_3$ (Figure 3) consists of eight bands, of which seven was successfully assigned using DFT+U framework while considering both selection rules and calculated absolute values of wavenumbers; we found out that the best match between theoretical and experimental wavenumbers is obtained if the former ones are scaled by the factor of 1.05 (Table 1 and Figure 3). The highest frequency band is placed at 448 cm$^{-1}$ and was assigned to B$_{2u}$ stretching within [AgF$_2$] planes in *ac* plane. The 377 cm$^{-1}$ shoulder cannot be assigned to any fundamental mode; it likely originates either from 126 cm$^{-1}$ + 252 cm$^{-1}$ combination mode (A$_{1u}$) or from an unknown impurity. In a lower frequency range, observed bands are rather broad and not fully resolved. One feature centred at 200 cm$^{-1}$, may be assigned to [AgF$_2$] plane buckling vibrations (B$_{2u}$).

Two shoulders at 295 and 252 cm$^{-1}$ are associated with similar plane buckling, as well as bending of the [AgF$^+$] chain (B$_{3u}$ and B$_{1u}$, respectively). The 136 cm$^{-1}$ band of medium strength is also assigned to chain bending (B$_{2u}$), but its two shoulders originate from potassium and fluoride lattice modes (both B$_{2u}$, at 110$^{-1}$ and 59 cm$^{-1}$, respectively).

In the measured room-temperature Raman spectrum seven bands are clearly visible. Two bands at 444 cm$^{-1}$ and 359 cm$^{-1}$ were assigned to Ag-F stretching within chain-forming [AgF$_6$] octahedra and within [AgF$_2$] planes (B$_{2g}$ and A$_g$, respectively). Weak band at 196 cm$^{-1}$ is associated with [AgF$_2$] intra sheet bending vibrations (B$_{2g}$). The lowest lying 126 cm$^{-1}$ and 101 cm$^{-1}$ features correspond, respectively, to B$_{1g}$ and/or a nearby A$_g$, and B$_{2g}$ lattice modes involving potassium cations. There are two more bands in the spectral region above 800 cm$^{-1}$ (not shown); the shoulder at 899 cm$^{-1}$ is supposedly the first overtone of the IR-active B$_{2u}$ mode at 448 cm$^{-1}$; the origin of the 949 cm$^{-1}$ band currently remains unknown. The assignment presented above is substantially strengthened by the fact that HSE06 hybrid DFT calculations yield identical band assignment without necessity to scale wavenumbers by the 1.05 factor.

The measured low-temperature Raman spectra are compared with the room-temperature one in Figure 4. Both spectra are characteristic of the same four broad main features, while in the low-temperature spectrum these features are shifted to slightly higher energies. All bands that are present at low temperature, are observable also at RT, and general shape of the spectrum remains unchanged. There is some phonon stiffening observed at low temperature, with the 444 cm$^{-1}$ band now appearing at 464 cm$^{-1}$, 359 cm$^{-1}$ one at 380 cm$^{-1}$, 196 cm$^{-1}$ one at 217 cm$^{-1}$, 126 cm$^{-1}$ one at 127 cm$^{-1}$ and 101 cm$^{-1}$ one at 109 cm$^{-1}$.

In published work, the disordered room temperature *Pcma* structure is derived by dividing *Pnma* unit cell in half, and introducing disorder of fluoride anions. The fluorine anions are placed alternately along the [AgF$^+$] chains with an occupancy of 0.5 [2]. As this type of phase transition does not substantially alter the network of covalent bonds, the thermal shift observed in Raman spectra can be satisfactorily explained by thermal compressibility of KAgF$_3$ lattice. With decreasing temperature Ag–F bond lengths also decrease, causing energy of associated stretching and bending phonons to rise. The fact that the room temperature vibrational spectra of KAgF$_3$ may be successfully assigned using the low-temperature *Pnma* structure serves as a confirmation of the X-ray diffraction-based hypothesis described above. This is additionally confirmed by comparing Raman spectrum obtained at room temperature and at 123 K (Figure 4, Table 2).

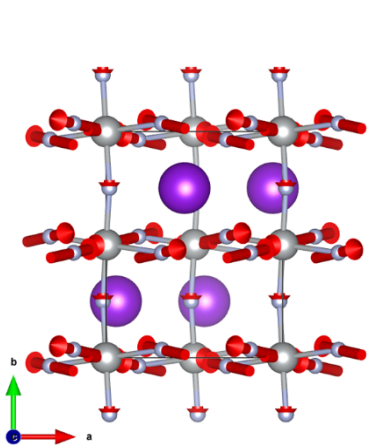

B$_{2u}$   Experimental: 448 cm$^{-1}$
105% DFT+U: 452 cm$^{-1}$
HSE06: 449 cm$^{-1}$

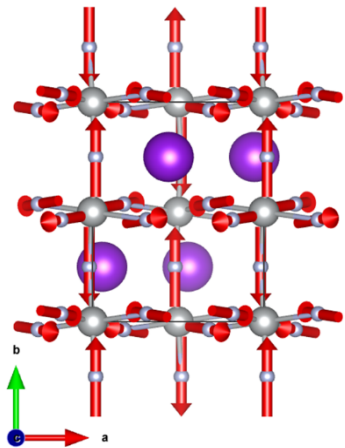

B$_{3g}$   Experimental: 444 cm$^{-1}$
105% DFT+U: 447 cm$^{-1}$
HSE06: 448 cm$^{-1}$

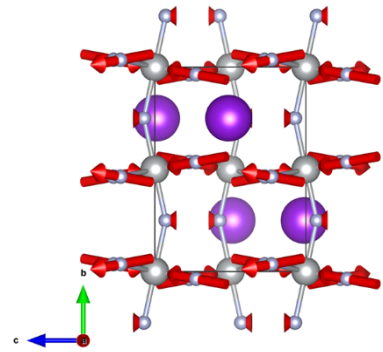

A$_g$   Experimental: 359 cm$^{-1}$
105% DFT+U: 367 cm$^{-1}$
HSE06: 377 cm$^{-1}$

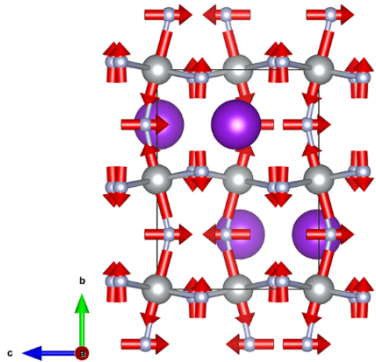

B$_{3u}$   Experimental: 295 cm$^{-1}$
105% DFT+U: 285 cm$^{-1}$
HSE06: 286 cm$^{-1}$

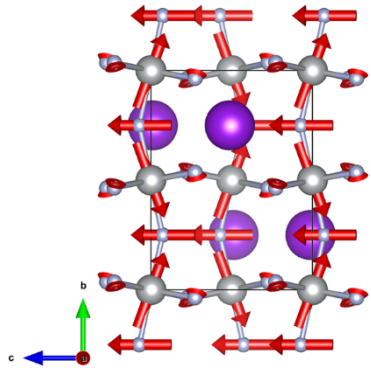

B$_{1u}$   Experimental: 252 cm$^{-1}$
105% DFT+U: 258 cm$^{-1}$
HSE06: 258 cm$^{-1}$

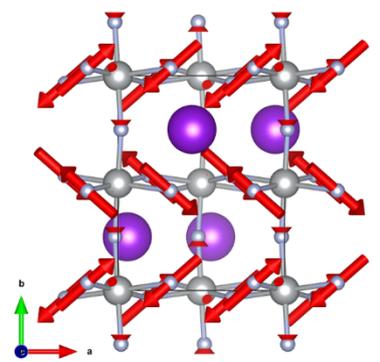

B$_{2u}$   Experimental: 200 cm$^{-1}$
105% DFT+U: 202 cm$^{-1}$
HSE06: 198 cm$^{-1}$

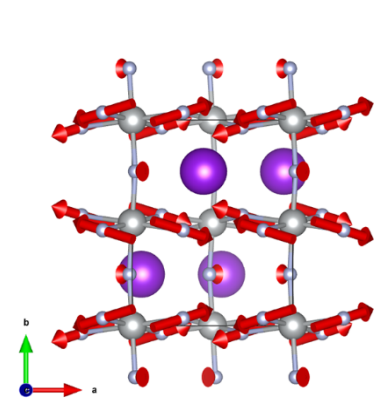

B$_{2g}$   Experimental: 196 cm$^{-1}$
105% DFT+U: 206 cm$^{-1}$
HSE06: 204 cm$^{-1}$

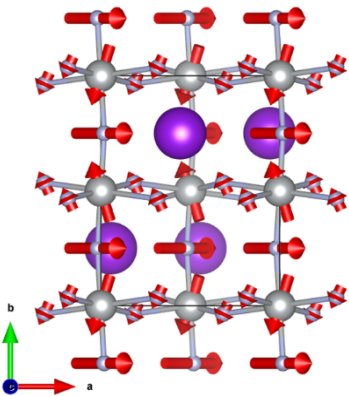

B$_{2u}$   Experimental: 136 cm$^{-1}$
105% DFT+U: 139 cm$^{-1}$
HSE06: 136 cm$^{-1}$

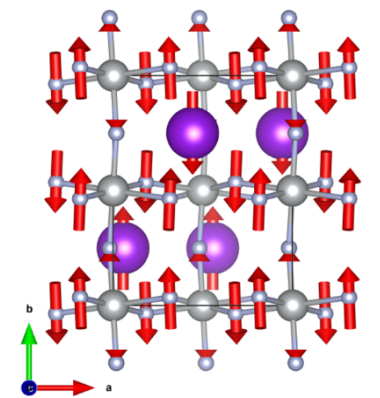

B$_{1g}$(A$_g$)   Experimental: 126 cm$^{-1}$
105% DFT+U: 126(123) cm$^{-1}$
HSE06: 127(121) cm$^{-1}$

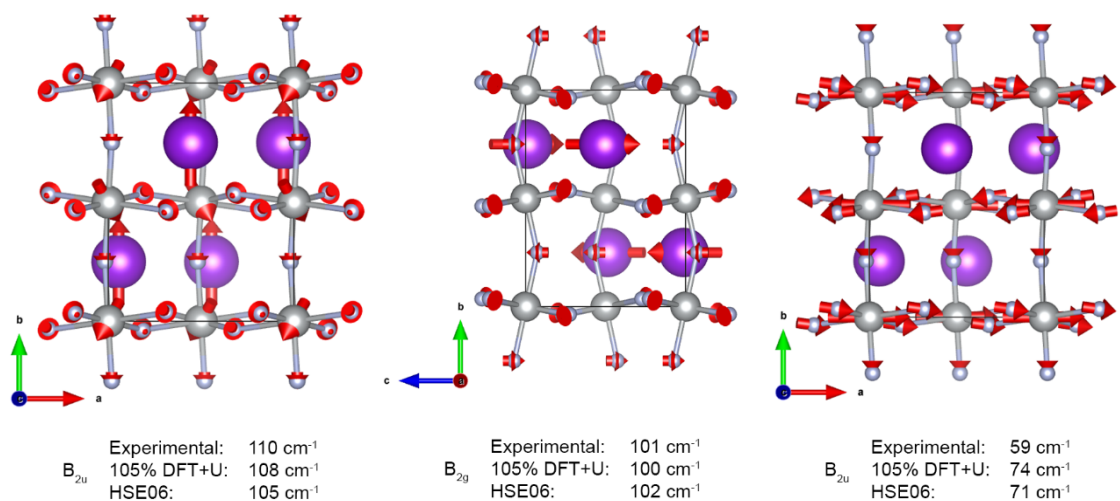

**Figure 6.** Normal vectors for selected fundamentals. Red arrows show the direction and relative amplitude of atomic movement

**Table 1.** Vibrational band assignment for KAgF$_3$.

| IR [cm$^{-1}$] | Raman [cm$^{-1}$] | DFT+U +5% [cm$^{-1}$] | HSE06 [cm$^{-1}$] | Symmetry | Assignment |
|---|---|---|---|---|---|
| | 949 m | – | | – | Impurity? |
| | 899 sh | – | | A$_g$ | Combination (2 × 448 cm$^{-1}$) |
| 448 vs | | 452 IR | 449 IR | B$_{2u}$ | [AgF$_2$] in plane stretching |
| | 444 vs | 447 R | 448 R | B$_{2g}$ | [AgF$_6$] octahedron stretching |
| 377 sh | | – | – | – | 126 cm$^{-1}$ + 252 cm$^{-1}$ combination mode (A$_{1u}$) or impurity |
| | 359 m | 367 R | 377 R | A$_g$ | [AgF$_2$] in plane stretching |
| 295 sh | | 285 IR | 286 IR | B$_{3u}$ | [AgF$_2$] plane buckling |
| 252 sh | | 258 IR | 258 IR | B$_{1u}$ | [AgF$^+$] chain bending |
| 200 s | | 202 IR | 198 IR | B$_{2u}$ | [AgF$_2$] plane buckling |
| | 196 w | 206 R | 204 R | B$_{2g}$ | [AgF$_2$] in plane bending |
| 136 m | | 139 IR | 136 IR | B$_{2u}$ | [AgF$^+$] chain bending |
| | 126[a] sh | 126 R | 127 R | B$_{1g}$ | [AgF$_2$] plane buckling + lattice (out of plane K) |
| | | 123 R | 121 R | A$_g$ | Lattice (in plane K) |
| 110 sh | | 108 IR | 105 IR | B$_{2u}$ | Lattice (out of plane K) |
| | 101 w | 100 R | 102 R | B$_{2g}$ | Lattice (in plane K) |
| 59 sh | | 74 IR | 71 IR | B$_{2u}$ | Lattice (in plane F) |

sh – shoulder, w – weak, m – medium, s – strong, vs – very strong, R – Raman active mode, IR – infrared active mode, [a] – mode with assignment to two nearby modes)

**Table 2.** Wavenumbers of selected Raman fundamentals [cm$^{-1}$] observed for KAgF$_3$ at 123 K and at room temperature, and their ratio, R [1].

| RT | 123 K | R (123 K/300K) |
|---|---|---|
| 444 vs | 465 vs | 1.05 |
| 359 m | 380 m | 1.06 |
| 196 w | 218 vw | 1.11 |
| 126 sh | 137 sh | 1.09 |
| 101 w | 106 vw | 1.05 |

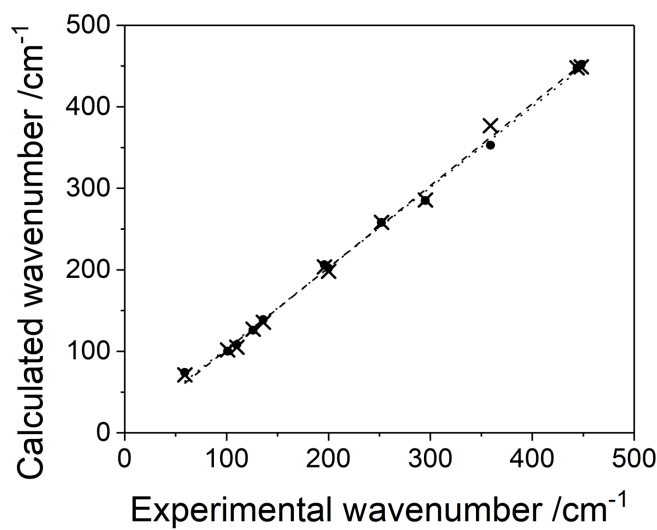

**Figure 7.** Correlations between theoretical and experimental wavenumbers jointly for IR and Raman spectra (black dotted line – linear regression with $R^2$=0.999 for corrected DFT+U dataset, black dashed line – linear regression with $R^2$=0.997 for HSE06 dataset; both not bound to pass at (0,0))

## 5. Conclusions

The vibrational spectra of KAgF$_3$ perovskite, a unique 1D antiferromagnet, have successfully been assigned with the help from theoretical DFT calculations. The calculated frequencies are in excellent agreement with experimental values when applying a 1.05 scaling factor for DFT+U calculations, and without any scaling for hybrid DFT ones. Phonon dispersion curves show that the low-temperature *Pnma* structure is dynamically stable, as no phonon softening is present. Comparison of low temperature and room temperature Raman measurements proves that order-disorder phase transition in 230 K has no major effect on the crystal structure and vibrational characteristics of the compound. To achieve complete understanding of the vibrational structure of KAgF$_3$ additional studies using e.g. Inelastic Neutron Scattering are needed which could determine the energies of these phonons, which have so far eluded experiment, and enable comparison with theoretical values.

**Supplementary Materials:** Table S1: List of all theoretically predicted normal modes, Figure S1: Visualization of theoretically predicted normal modes in order of decreasing wavenumber.

**Author Contributions:** For research articles with several authors, a short paragraph specifying their individual contributions must be provided. The following statements should be used "Conceptualization, W.G. and M.D.; methodology, Z.M. and M.D.; formal analysis, K.K.; investigation, K.K., J.G. and Z.M.; resources, W.G.; writing—original draft preparation, K.K.; writing—review and editing, M.D. and W.G.; visualization, K.K.; supervision, M.D. and W.G.; project administration, W.G.; funding acquisition, W.G. All authors have read and agreed to the published version of the manuscript."

**Funding:** This work was supported by Polish National Science Center (NCN) within Maestro project (2017/26/A/ST5/00570). funding. Dr Derzsi acknowledges the ERDF, R&I Operational Program (ITMS2014+: 313011W085), Scientific Grant Agency of the Slovak Republic grant (VG 1/0223/19) and the Slovak Research and Development Agency grant (APVV-18-0168).

**Acknowledgments:** The research was carried out using supercomputers of Interdisciplinary Centre for Mathematical and Computational Modelling (ICM), University of Warsaw, under grant number GA76-19 (ADVANCE++). Experimental research in Warsaw was carried out with the use of CePT infrastructure financed by the European Union - the European Regional Development Fund within the Operational Programme "Innovative economy" for 2007-2013 (POIG.02.02.00-14-024/08-00).

**Conflicts of Interest:** The authors declare no conflict of interest.

**References**

1. Mazej, Z.; Goreshnik, E.; Jagliić, Z.; Gaweł, B.; Łasocha, W.; Grzybowska, D.; Jaroń, T.; Kurzydłowski, D.; Malinowski, P.; Koźminski, W.; Szydłowska, J.; Leszczyński, P.; Grochala, W. $KAgF_3$, $K_2AgF_4$ and $K_3Ag_2F_7$: Important steps towards a layered antiferromagnetic fluoroargentate(II). *CrystEngComm* **2009**, *11*, 1702–1710. https://doi.org/10.1039/b902161b.
2. Kurzydłowski, D.; Mazej, Z.; Jagličić, Z.; Filinchuk, Y.; Grochala W. Structural transition and unusually strong antiferromagnetic superexchange coupling in perovskite $KAgF_3$. *Chem. Commun.* **2013**, *49*, 6262–6264. https://doi.org/10.1039/c3cc41521j.
3. Zhang, X.; Zhang, G.; Jia, T.; Guo, Y.; Zeng, Z.; Lin, H.Q. KAgF3: Quasi-one-dimensional magnetism in three-dimensional magnetic ion sublattice. *Phys. Lett. A* **2011**, *375*, 2456–2459. https://doi.org/10.1016/j.physleta.2011.05.025.
4. Kurzydłowski, D.; Grochala, W. Large exchange anisotropy in quasi-one-dimensional spin-1/2 fluoride antiferromagnets with a $d(z^2)^1$ ground state. *Phys. Rev. B* **2017**, *96*, 155140. https://doi.org/10.1103/PhysRevB.96.155140.
5. Derzsi, M.; Malinowski, P.J.; Mazej, Z.; Grochala, W. Phonon spectra and phonon-dependent properties of $AgSO_4$, an unusual sulfate of divalent silver. *Vib. Spectrosc.* **2011**, *57*, 334–337. https://doi.org/10.1016/j.vibspec.2011.07.001.
6. Gawraczyński, J.; Kurzydłowski, D.; Gadomski, W.; Mazej, Z.; Ruani, G.; Bergenti, I.; Jaroń, T.; Ozarowski, A.; Hill, S.; Leszczyński, P.J.; Tokár, K.; Derzsi, M.; Barone, P.; Wohlfeld, K.; Lorenzana, J.; Grochala W. The silver route to cuprate analogs. *Proc. Natl. Acad. Sci. U. S. A.* **2019**, *116*, 1495–1500. https://doi.org/https://doi.org/10.1073/pnas.1812857116.
7. Liechtenstein, A.I.; Anisimov, V.I.; Zaanen, J. Density-functional theory and strong interactions: Orbital ordering in Mott-Hubbard insulators. *Phys. Rev. B.* **1995**, *52*, R5467–R5470. https://doi.org/10.1103/PhysRevB.52.R5467.
8. Csonka, G.I.; Perdew, J.P.; Ruzsinszky, A.; Philipsen, P.H.T.; Lebègue, S.; Paier, J.; Vydrov, O.A.; Ángyán, J.G. Assessing the performance of recent density functionals for bulk solids. *Phys. Rev. B* **2009**, *79*, 155107. https://doi.org/10.1103/PhysRevB.79.155107.
9. Blöchl, P.E. Projector augmented-wave method. *Phys. Rev. B* **1994**, *50*, 17953–17979. https://doi.org/10.1103/PhysRevB.50.17953.
10. Kresse, G.; Joubert, D. From ultrasoft pseudopotentials to the projector augmented-wave method. *Phys. Rev. B* **1999**, *59*, 1758–1775. https://doi.org/10.1103/PhysRevB.59.1758.
11. Kresse, G.; Hafner, J. Ab initio molecular dynamics for liquid metals. *Phys. Rev. B* **1993**, *47*, 558–561. https://doi.org/10.1103/PhysRevB.47.558.
12. Kresse, G.; Furthmüller, J. Efficient iterative schemes for ab initio total-energy calculations using a plane-wave basis set. *Phys. Rev. B* **1996**, *54*, 11169–11186. https://doi.org/10.1103/PhysRevB.54.11169.
13. Kresse, G.; Furthmüller, J. Efficiency of ab-initio total energy calculations for metals and semiconductors using a plane-wave basis set. *Comput. Mater. Sci.* **1996**, *6*, 15–50. https://doi.org/10.1016/0927-0256(96)00008-0.
14. Parlinski, K.; Li, Z.Q.; Kawazoe, Y. First-principles determination of the soft mode in cubic ZrO2. *Phys. Rev. Lett.* **1997**, *78*, 4063–4066. https://doi.org/10.1103/PhysRevLett.78.4063.
15. Parlinski, K.; PHONON Software **2017**.
16. Krukau, A. V.; Vydrov, O.A.; Izmaylov, A.F.; Scuseria, G.E. Influence of the exchange screening parameter on the performance of screened hybrid functionals. *J. Chem. Phys.* **2006**, *125*, 224106. https://doi.org/10.1063/1.2404663.

*Supplementary material*

**Table S1.** List of all theoretically predicted normal modes (modes assigned to observable bands have been marked in bold fonts and grey background).

| No. | 105% DFT+U [cm$^{-1}$] | HSE06 [cm$^{-1}$] | Irreducible representation |
|---|---|---|---|
| 1 | 469 | 460 | B2u |
| 2 | 467 | 458 | Au |
| 3 | 458 | 456 | B1u |
| 4 | 457 | 455 | B3u |
| 5 | 454 | 454 | Au |
| **6** | **452** | **449** | **B2u** |
| **7** | **447** | **448** | **B3g** |
| 8 | 428 | 441 | B2g |
| 9 | 406 | 419 | B1g |
| **10** | **367** | **377** | **Ag** |
| 11 | 353 | 356 | B1g |
| 12 | 292 | 301 | B3g |
| **13** | **285** | **286** | **B3u** |
| **14** | **258** | **258** | **B1u** |
| 15 | 229 | 229 | B3u |
| 16 | 228 | 226 | B2g |
| 17 | 227 | 224 | Ag |
| 18 | 221 | 220 | B1u |
| 19 | 212 | 214 | B1u |
| **20** | **206** | **204** | **B2g** |
| 21 | 204 | 201 | Au |
| **22** | **202** | **198** | **B2u** |
| 23 | 193 | 194 | B3u |
| 24 | 181 | 178 | B3g |
| 25 | 174 | 171 | B2u |
| 26 | 174 | 170 | B1g |
| 27 | 173 | 169 | Au |
| 28 | 164 | 164 | B3u |
| 29 | 162 | 162 | Ag |
| 30 | 162 | 162 | B1u |
| 31 | 158 | 154 | B3u |
| 32 | 155 | 152 | B1u |
| 33 | 148 | 146 | B3g |
| 34 | 147 | 145 | B2g |
| 35 | 142 | 138 | B1u |
| 36 | 139 | 137 | B3u |
| **37** | **139** | **136** | **B2u** |
| 38 | 136 | 133 | Au |
| 39 | 135 | 133 | B2g |
| 40 | 134 | 130 | Ag |
| **41** | **126** | **127** | **B1g** |
| 42 | 123 | 123 | B3g |
| **43** | **123** | **121** | **Ag** |
| 44 | 121 | 119 | B1u |
| 45 | 121 | 118 | B3u |
| **46** | **108** | **105** | **B2u** |
| 47 | 104 | 103 | Au |
| 48 | 100 | 102 | B2g |
| **49** | **97** | **99** | **B1g** |
| 50 | 93 | 97 | B2g |
| 51 | 88 | 91 | B1u |
| 52 | 87 | 86 | Ag |
| 53 | 79 | 81 | Ag |
| 54 | 76 | 73 | B3u |
| **55** | **74** | **71** | **B2u** |
| 56 | 72 | 70 | Au |
| 57 | 64 | 61 | Au |
| 58 | 0 | 1 | B3u |
| 59 | 0 | 1 | B2u |
| 60 | 0 | 2 | B1u |

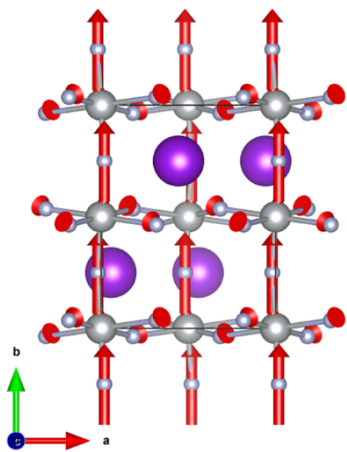
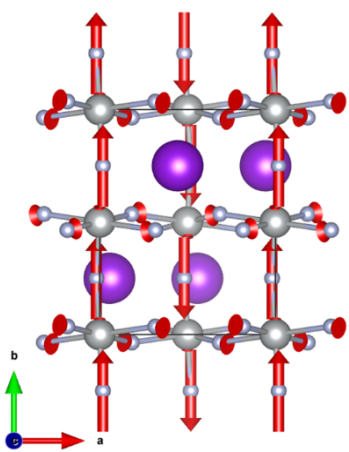
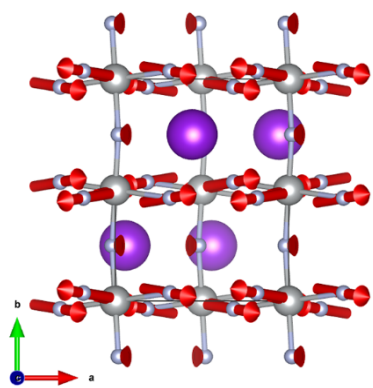

105% DFT+U:   469 cm⁻¹          105% DFT+U:   467 cm⁻¹          105% DFT+U:   458 cm⁻¹
HSE06:        460 cm⁻¹          HSE06:        458 cm⁻¹          HSE06:        456 cm⁻¹

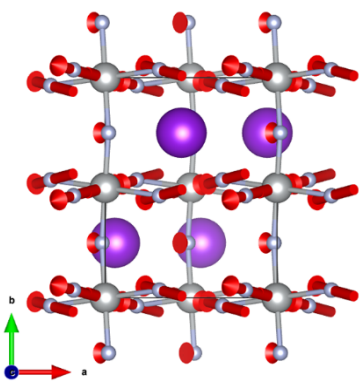
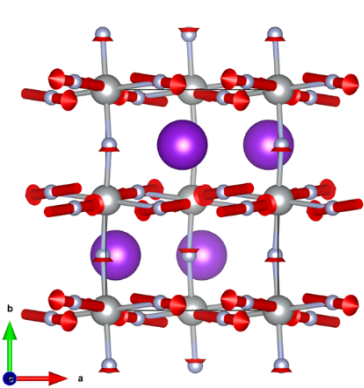
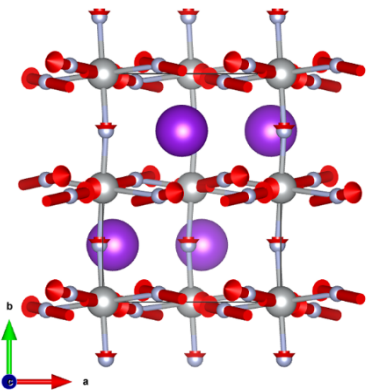

105% DFT+U:   457 cm⁻¹          105% DFT+U:   454 cm⁻¹          105% DFT+U:   452 cm⁻¹
HSE06:        455 cm⁻¹          HSE06:        454 cm⁻¹          HSE06:        449 cm⁻¹

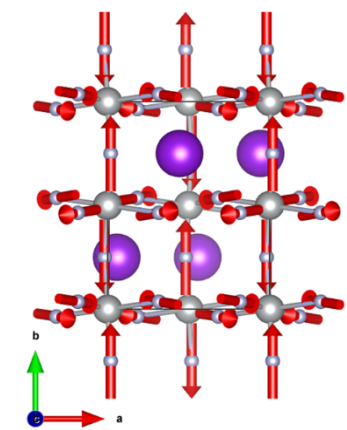
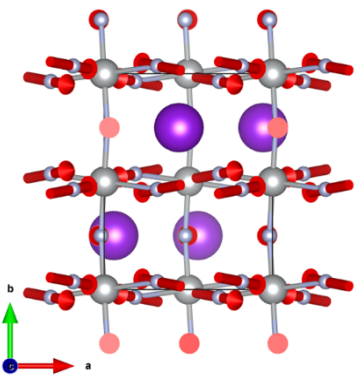
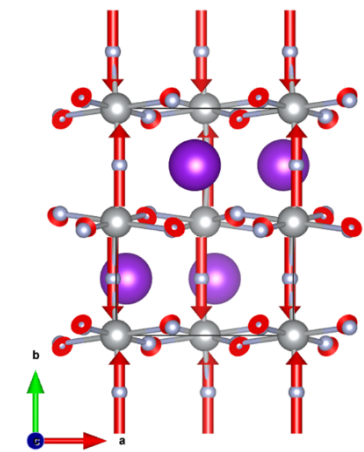

105% DFT+U:   447 cm⁻¹          105% DFT+U:   428 cm⁻¹          105% DFT+U:   406 cm⁻¹
HSE06:        448 cm⁻¹          HSE06:        441 cm⁻¹          HSE06:        419 cm⁻¹

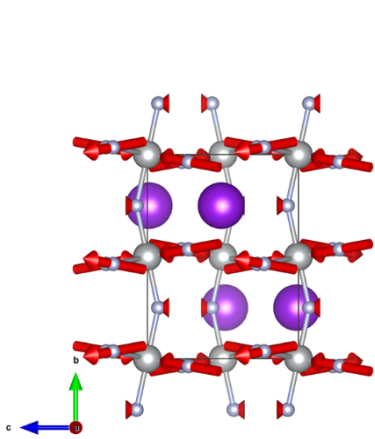

105% DFT+U:  367 cm⁻¹
HSE06:       377 cm⁻¹

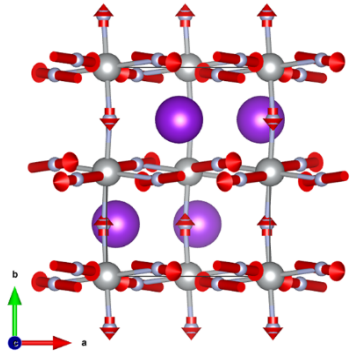

105% DFT+U:  353 cm⁻¹
HSE06:       356 cm⁻¹

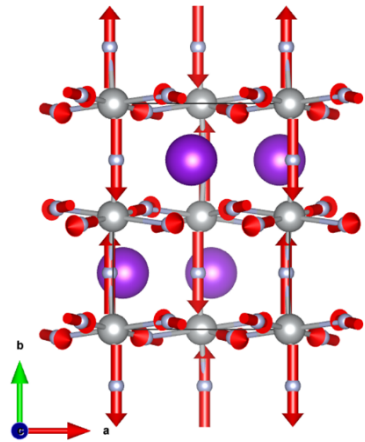

105% DFT+U:  292 cm⁻¹
HSE06:       301 cm⁻¹

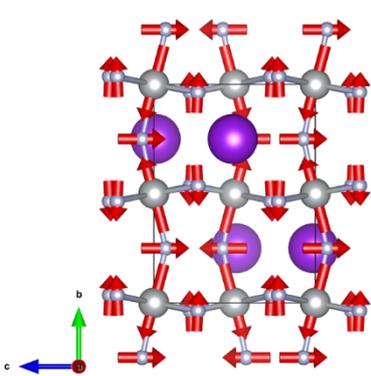

105% DFT+U:  285 cm⁻¹
HSE06:       286 cm⁻¹

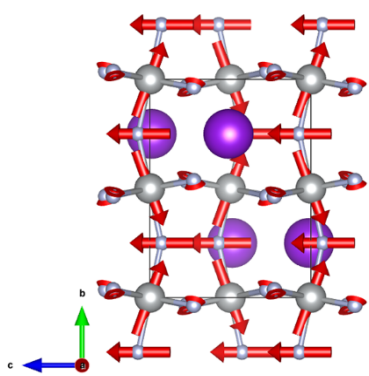

105% DFT+U:  258 cm⁻¹
HSE06:       258 cm⁻¹

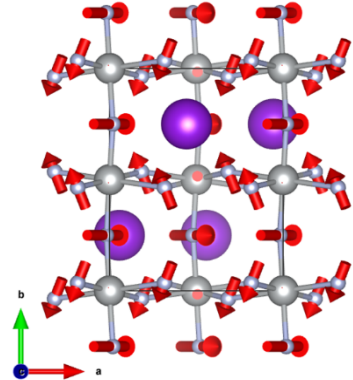

105% DFT+U:  229 cm⁻¹
HSE06:       229 cm⁻¹

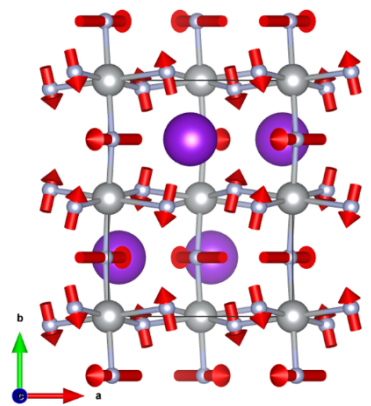

105% DFT+U:  228 cm⁻¹
HSE06:       226 cm⁻¹

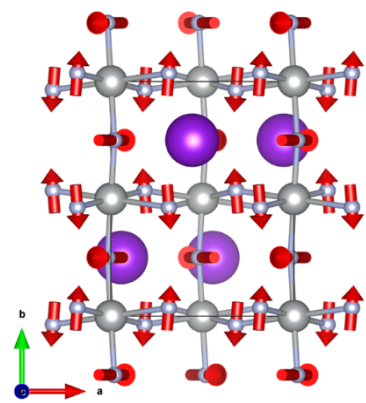

105% DFT+U:  227 cm⁻¹
HSE06:       224 cm⁻¹

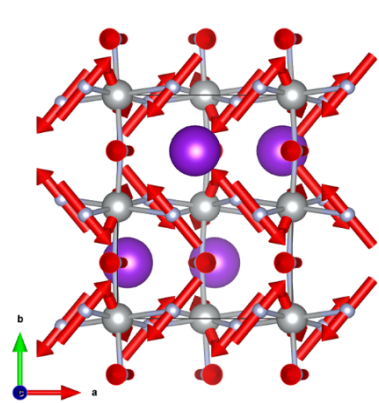

105% DFT+U:  221 cm⁻¹
HSE06:       220 cm⁻¹

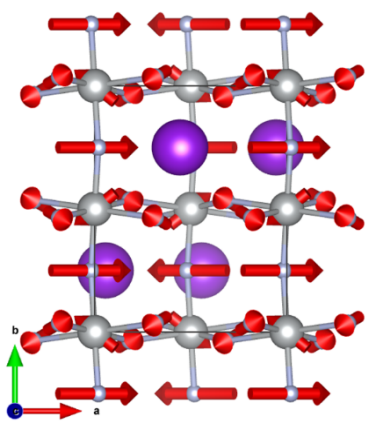
105% DFT+U:  212 cm⁻¹
HSE06:       214 cm⁻¹

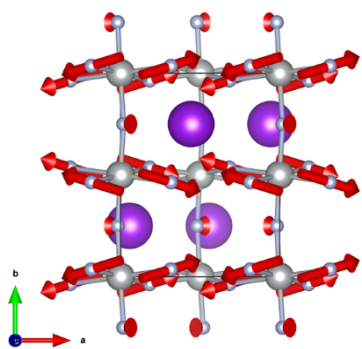
105% DFT+U:  206 cm⁻¹
HSE06:       204 cm⁻¹

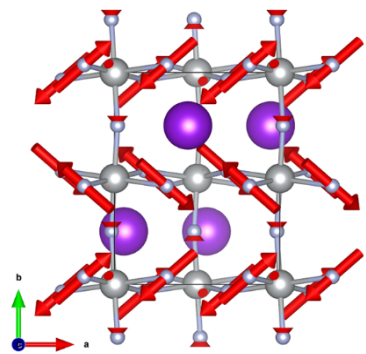
105% DFT+U:  204 cm⁻¹
HSE06:       201 cm⁻¹

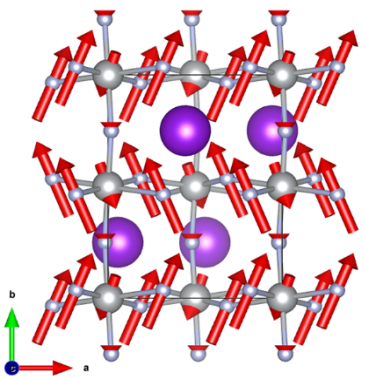
105% DFT+U:  201 cm⁻¹
HSE06:       198 cm⁻¹

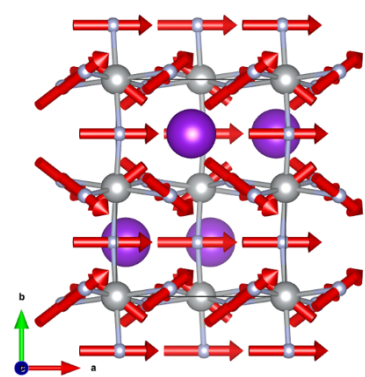
105% DFT+U:  193 cm⁻¹
HSE06:       194 cm⁻¹

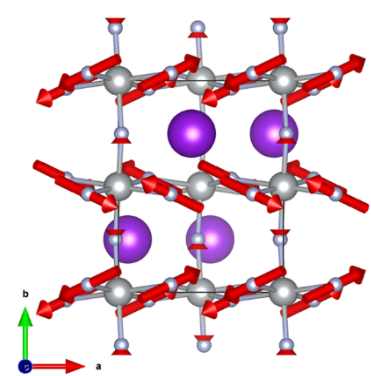
105% DFT+U:  181 cm⁻¹
HSE06:       178 cm⁻¹

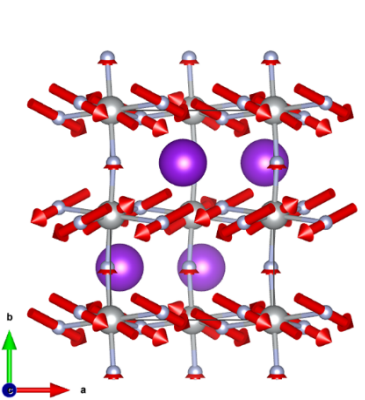
105% DFT+U:  174 cm⁻¹
HSE06:       171 cm⁻¹

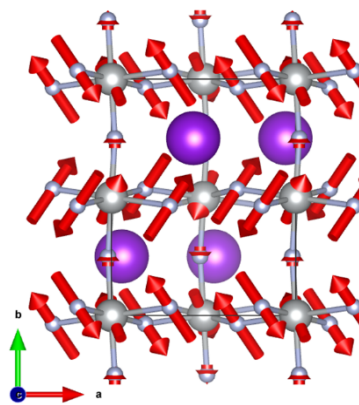
105% DFT+U:  174 cm⁻¹
HSE06:       170 cm⁻¹

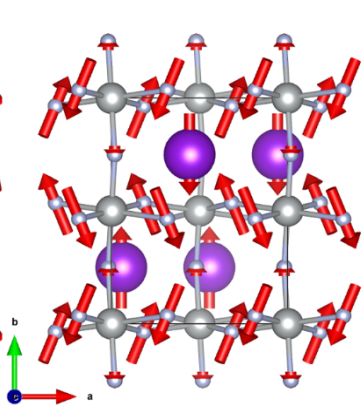
105% DFT+U:  174 cm⁻¹
HSE06:       169 cm⁻¹

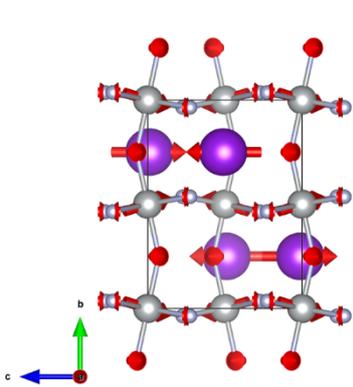 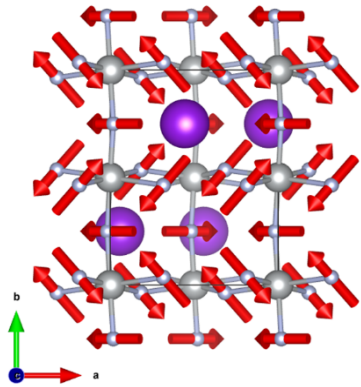 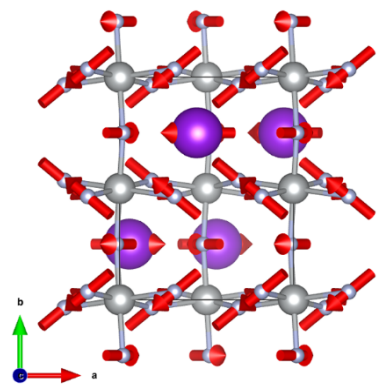

105% DFT+U:  164 cm⁻¹  105% DFT+U:  162 cm⁻¹  105% DFT+U:  162 cm⁻¹
HSE06:       164 cm⁻¹  HSE06:       162 cm⁻¹  HSE06:       162 cm⁻¹

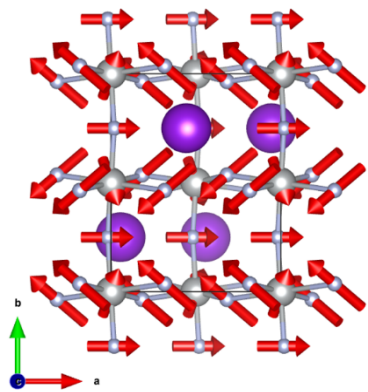 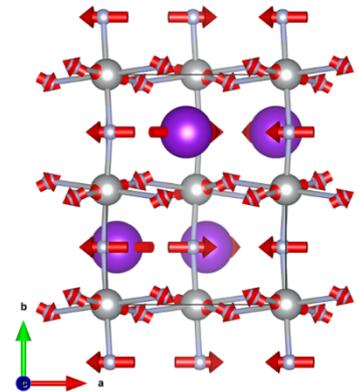 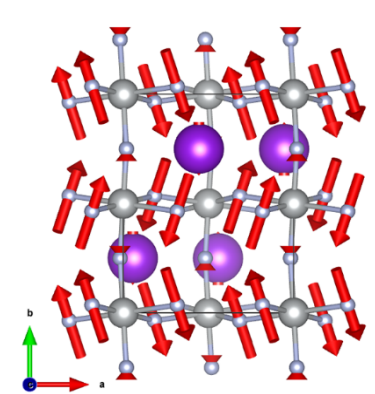

105% DFT+U:  158 cm⁻¹  105% DFT+U:  155 cm⁻¹  105% DFT+U:  148 cm⁻¹
HSE06:       154 cm⁻¹  HSE06:       152 cm⁻¹  HSE06:       146 cm⁻¹

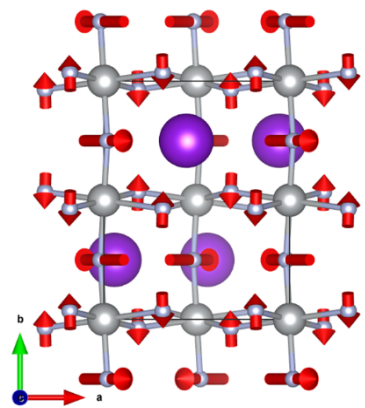 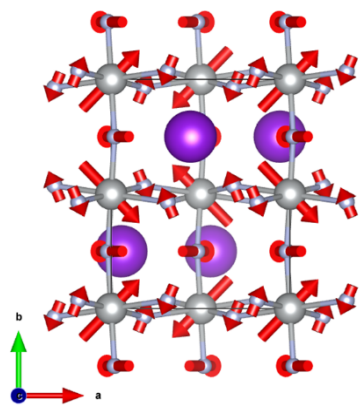 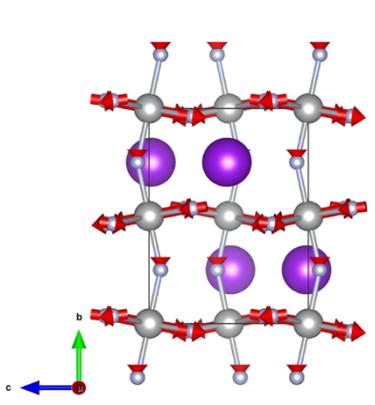

105% DFT+U:  147 cm⁻¹  105% DFT+U:  142 cm⁻¹  105% DFT+U:  139 cm⁻¹
HSE06:       145 cm⁻¹  HSE06:       138 cm⁻¹  HSE06:       137 cm⁻¹

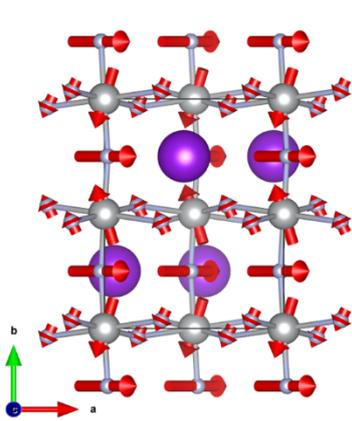 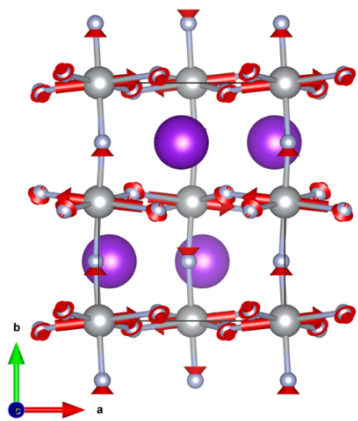 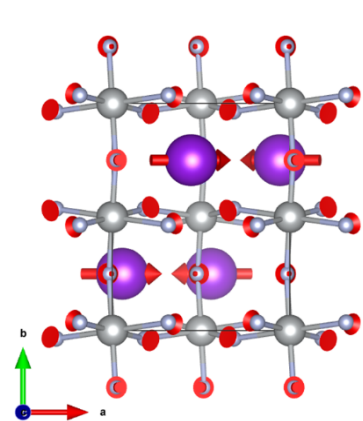

| | | |
|---|---|---|
| 105% DFT+U: 139 cm$^{-1}$ | 105% DFT+U: 136 cm$^{-1}$ | 105% DFT+U: 135 cm$^{-1}$ |
| HSE06: 136 cm$^{-1}$ | HSE06: 133 cm$^{-1}$ | HSE06: 133 cm$^{-1}$ |

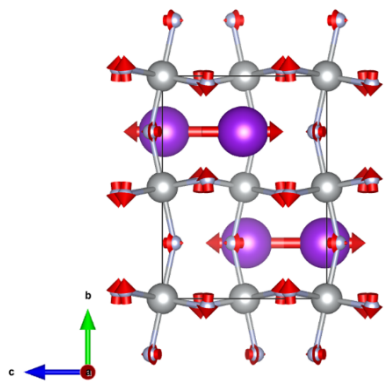 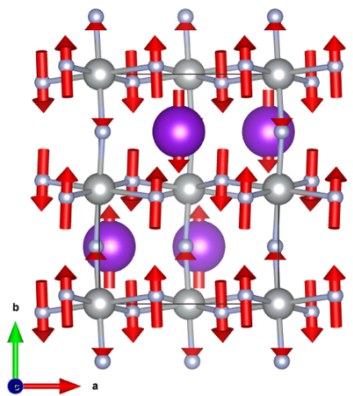 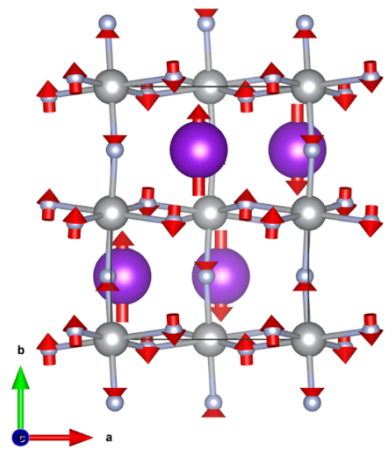

| | | |
|---|---|---|
| 105% DFT+U: 134 cm$^{-1}$ | 105% DFT+U: 126 cm$^{-1}$ | 105% DFT+U: 123 cm$^{-1}$ |
| HSE06: 130 cm$^{-1}$ | HSE06: 127 cm$^{-1}$ | HSE06: 123 cm$^{-1}$ |

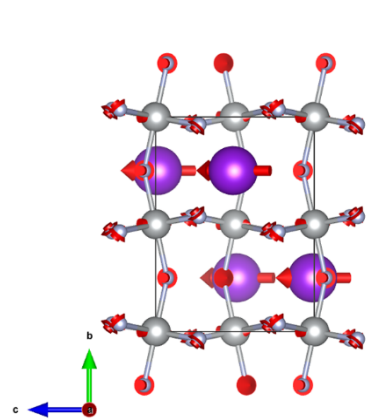 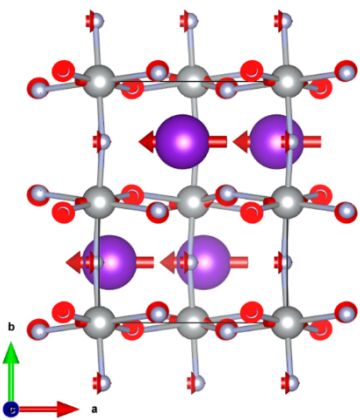 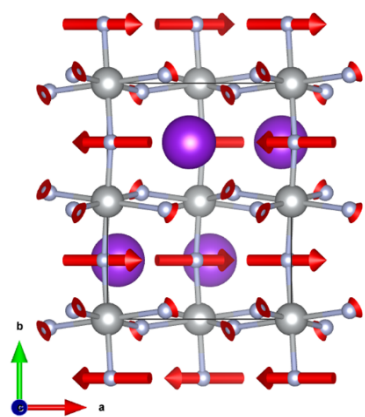

| | | |
|---|---|---|
| 105% DFT+U: 123 cm$^{-1}$ | 105% DFT+U: 121 cm$^{-1}$ | 105% DFT+U: 121 cm$^{-1}$ |
| HSE06: 121 cm$^{-1}$ | HSE06: 119 cm$^{-1}$ | HSE06: 118 cm$^{-1}$ |

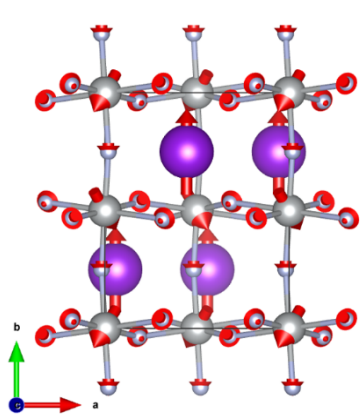
105% DFT+U:   108 cm$^{-1}$
HSE06:            105 cm$^{-1}$

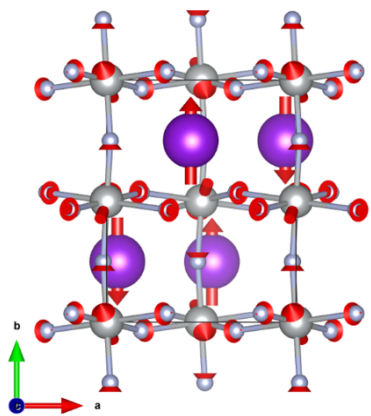
105% DFT+U:   104 cm$^{-1}$
HSE06:            103 cm$^{-1}$

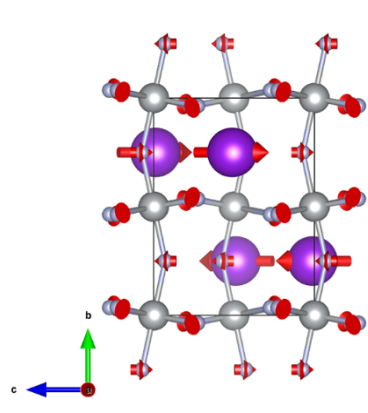
105% DFT+U:   100 cm$^{-1}$
HSE06:            102 cm$^{-1}$

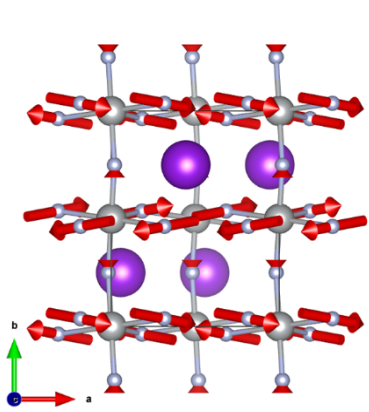
105% DFT+U:   97 cm$^{-1}$
HSE06:            99 cm$^{-1}$

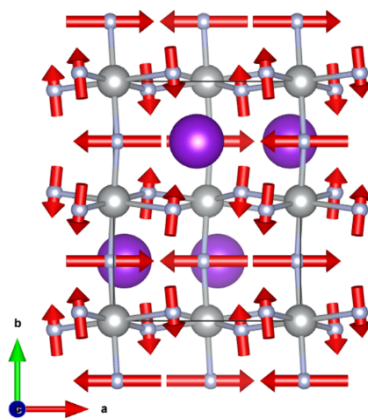
105% DFT+U:   93 cm$^{-1}$
HSE06:            97 cm$^{-1}$

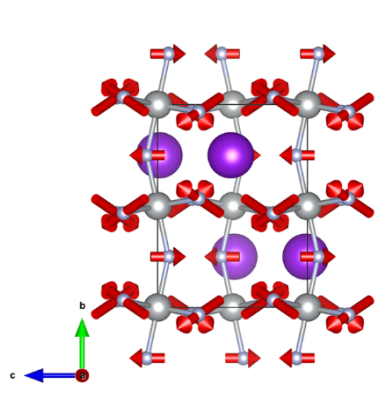
105% DFT+U:   88 cm$^{-1}$
HSE06:            91 cm$^{-1}$

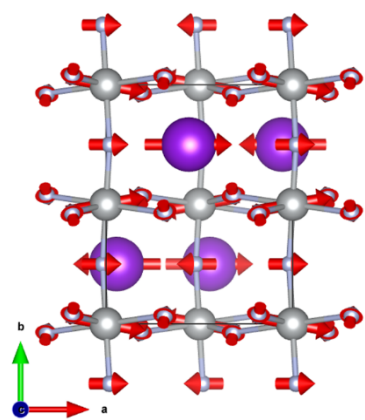
105% DFT+U:   87 cm$^{-1}$
HSE06:            86 cm$^{-1}$

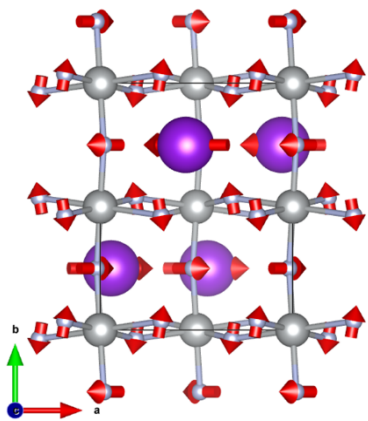
105% DFT+U:   79 cm$^{-1}$
HSE06:            81 cm$^{-1}$

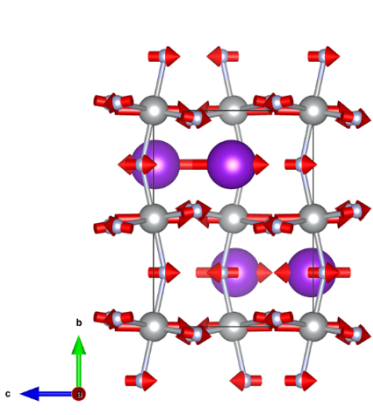
105% DFT+U:   76 cm$^{-1}$
HSE06:            73 cm$^{-1}$

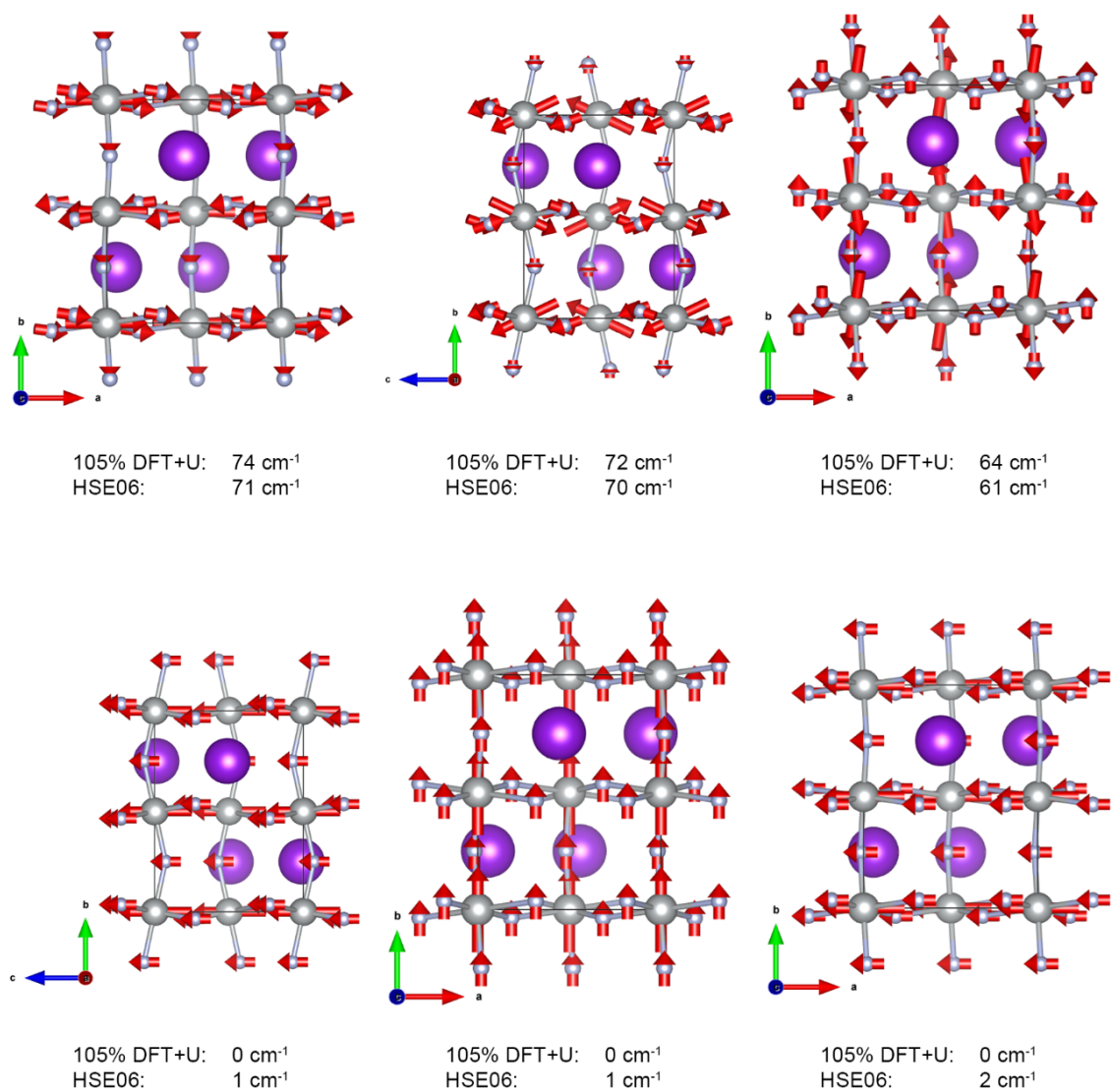

**Figure S1.** Visualization of theoretically predicted normal modes in order of decreasing wavenumbers